\documentclass[aps,prb,10pt,twocolumn,superscriptaddress]{revtex4-1}

\usepackage{amsmath, amssymb}    					
\usepackage{graphicx}   								
\usepackage{xcolor}     								
\usepackage{hyperref}   								
\hypersetup{colorlinks=true,allcolors=blue}			
\usepackage{textcomp}								

\newcommand{\ddt}[0]{\frac{\partial}{\partial t}}
\renewcommand{\t}[1]{\textrm{#1}}
\newcommand{\nn}[0]{\nonumber\\}

\newcommand{\q}[0]{\mathbf{q}}

\newcommand{\up}[0]{\uparrow}
\newcommand{\down}[0]{\downarrow}
\newcommand{\Jsd}[0]{J_{sd}}
\newcommand{\Jpd}[0]{J_{pd}}
\newcommand{\NMn}[0]{N_\t{Mn}}

\newcommand{\etah}{\eta_{\t{h}}}

\newcommand{\ud}{{\uparrow/\downarrow}}
\newcommand{\du}{{\downarrow/\uparrow}}

\begin{document}

\title{Origins of overshoots in the exciton spin dynamics in semiconductors}
\author{F. Ungar}
\affiliation{Theoretische Physik III, Universit\"at Bayreuth, 95440 Bayreuth, Germany}
\author{M. Cygorek}
\affiliation{Department of Physics, University of Ottawa, Ottawa, Ontario, Canada K1N 6N5}
\author{V. M. Axt}
\affiliation{Theoretische Physik III, Universit\"at Bayreuth, 95440 Bayreuth, Germany}

\begin{abstract}

We investigate the origin of overshoots in the exciton spin dynamics after resonant optical excitation.
As a material system, we focus on diluted magnetic semiconductor quantum wells as they provide a strong spin-flip scattering for the carriers.
Our study shows that overshoots can appear as a consequence of radiative decay even on the single-particle level in a theory without any memory.
The magnitude of the overshoots in this case depends strongly on the temperature as well as the doping fraction of the material.
If many-body effects beyond the single-particle level become important so that a quantum-kinetic description is required, a spin overshoot appears already without radiative decay and is much more robust against variations of system parameters.
We show that the origin of the spin overshoot can be determined either via its temperature dependence or via its behavior for different doping fractions.
The results can be expected to apply to a wide range of semiconductors as long as radiative decay and a spin-flip mechanism are present.

\end{abstract}

\maketitle

\section{Introduction}
\label{sec:Introduction}

It is well known that, whereas the dynamics of isolated few-level quantum systems is oscillatory with frequencies corresponding to its eigenenergies, quantum systems weakly coupled to a Markovian environment exhibit an exponentially decaying dynamics.
If, however, the coupling to the bath is not weak or the bath is
non-Markovian, the decay is, in general, not exponential. Instead,
traces of the underlying coherent oscillatory behavior can remain visible in the dynamics in the form of overshoots.
Thus, signal overshoots are a quite fundamental property of many physical systems.
In the literature, overshoots are encountered in a wide variety of material systems and have recently been discussed and observed, e.g., in vertical-cavity surface-emitting lasers \cite{Mulet_Current-self_2008}, the ultrafast dynamics of amorphous magnets \cite{Becker_Ultrafast-laser_2015}, or in the spin-lattice relaxation measured via nuclear magnetic resonance \cite{Fu_Revisiting-spin_2016}.

Here, we focus on the latter area of spin physics and provide a theoretical description of spin overshoots in diluted magnetic semiconductors (DMSs), a material class where standard semiconductors are doped with a small number of magnetic impurities such as manganese \cite{Furdyna_Diluted-magnetic_1988, Furdyna_Semiconductor-and_1988, Kossut_Introduction-to_2010, Dietl_Dilute-ferromagnetic_2014}.
Besides their importance for possible spintronics applications \cite{Zutic_Spintronics-Fundamentals_2004, Awschalom_Challenges-for_2007, Awschalom_Quantum-Spintronics_2013, Dietl_A-ten_2010, Ohno_A-window_2010, Joshi_Spintronics_2016}, these materials exhibit rich many-body physics \cite{Ohno_Making-Nonmagnetic_1998, DiMarco_Electron-correlations_2013, Ungar_Many-body_2018} due to pronounced correlation effects between the carrier and the impurity subsystem.

In II-VI DMSs, where the doping with Mn ions does not lead to additional charges, the impurities provide a strong spin-flip mechanism for the carriers via the exchange interaction which typically dominates the spin dynamics \cite{Dietl_Dilute-ferromagnetic_2014, Kossut_Introduction-to_2010}.
Recent theoretical work shows that an adequate description of the spin dynamics in these systems actually requires a treatment of the exchange interaction beyond the single-particle level in order to account for carrier-impurity correlations \cite{Thurn_Non-Markovian_2013, Cygorek_Non-Markovian_2015, Ungar_Quantum-kinetic_2017, Ungar_Trend-reversal_2018}.
Using such a quantum kinetic approach, the finite memory induced by the correlations was found to lead to overshoots in the spin dynamics both for quasi-free carriers \cite{Thurn_Non-Markovian_2013} as well as for electron-hole pairs bound by the Coulomb interaction \cite{Ungar_Quantum-kinetic_2017}.
Furthermore, it has been shown that for excitons such non-Markovian effects can even explain the quantitative deviation between spin-transfer rates obtained by Fermi's golden rule and the experimental data obtained by several independent groups \cite{BenCheikh_Electron-spin_2013, Ungar_Quantum-kinetic_2017}.

In general, spin overshoots represent a very attractive qualitative signature of non-Markovian effects since they are easy to distinguish from a monoexponential decay as predicted by Fermi's golden rule.
However, it turns out that overshoots can also be caused by another, much simpler mechanism:
the combination of radiative decay with optical spin selection rules.
To see this, one can envision a system with two spin channels where only one couples to the light field and the other is dark.
Then the occupation in the bright channel will decay while spins in the dark channel are not affected until a spin flip occurs.
Thus, when looking at the total spin given by the sum of the populations of the two channels, an overshoot can occur.
This appears very naturally already on the Markov level where no correlations are accounted for since the only requirements for such a dynamics are the existence of a bright and a dark spin channel as well as a suitably strong spin-flip mechanism.

It is therefore an important task to discern overshoots in the spin dynamics
in DMSs caused by radiative decay from those caused by genuine non-Markovian effects, which is the main goal of this paper.
To this end, simulations are performed for manganese-doped ZnSe quantum well nanostructures which are optically excited at the $1s$ exciton resonance.
On the Markov level, the origin of the spin overshoot can also be made quite transparent by going over to a minimal model, retaining only a decay of each spin population as well as a spin-transfer rate between them.
It is found that overshoots in this model caused by radiative decay are most pronounced at low temperatures where phonon absorption is negligible and that phonons significantly inhibit overshoots at elevated temperatures.
However, a quantum kinetic treatment of the exciton-impurity exchange interaction yields an overshoot which is much more robust against variations of the temperature and does not rely on a finite radiative decay rate.

In order to provide suggestions as to how the origin of spin overshoots can be determined in experiments for a particular sample, we provide a comparison of the results of the Markovian theory (MT) as well as the quantum kinetic theory (QKT) that reveals drastically different trends and dependences of the overshoot on parameters such as the temperature and the doping fraction.
These results can be expected to hold also for a larger class of materials since they in principle only rely on radiative decay and the presence of a spin-flip mechanism.
Thus, our investigation allows one to determine whether an overshoot encountered in the spin dynamics is dominated by radiative decay or many-body effects in a particular sample.

\section{Theoretical background}
\label{sec:Theoretical-background}

First, the constituting parts of the Hamiltonian which is used for the description of the spin dynamics in DMSs are briefly discussed and an intuitive explanation of the relevant spin-flip processes is given.
Furthermore, we provide the equation of motion for the spin-up and spin-down exciton density in the Markov approximation from which the spin dynamics can be calculated.
Finally, the general structure of this equation is discussed using a minimal model which also allows us to analyze the preconditions for the appearance of spin overshoots.

\subsection{Exciton-spin dynamics in diluted magnetic semiconductors}
\label{subsec:Exciton-spin-dynamics-in-diluted-magnetic-semiconductors}

Concerning the spin dynamics in DMSs, the dominant interaction is given by the $sd$ ($pd$) exchange interaction that induces spin flips between $s$-type electrons ($p$-type holes) and the localized $d$-shell electrons of the magnetic impurities \cite{Kossut_Introduction-to_2010}.
Apart from the magnetic exchange interaction, we also take the scattering of carriers with longitudinal acoustic (LA) phonons into account \cite{Ungar_Phonon-impact_2018, Ungar_Phonon-induced_2019}.
Longitudinal optical (LO) phonon scattering can be disregarded here since it is negligible below temperatures of about $80\,$K and the kinetic energies encountered here are well below the LO-phonon threshold \cite{Rudin_Temperature-dependent_1990}.
We also assume a linear phonon dispersion $\omega_\q^\t{ph} = vq$ because of the small exciton center-of-mass momenta involved in the dynamical processes after an optical excitation resonant with the exciton ground state.
Additional contributions to the Hamiltonian include the carrier kinetic energies, the Coulomb interaction responsible for the exciton binding, and the light-matter coupling in the dipole approximation.
Furthermore, the local potential mismatch in the lattice due to the doping with impurities is captured by adding a nonmagnetic scattering contribution that does not depend on the carrier spin \cite{Cygorek_Influence-of_2017}.
The model is further extended to include the radiative decay of excitons quantified by a fixed rate $\Gamma_0$ \cite{Ungar_Many-body_2018}.
Instead of providing the lengthy explicit expressions for all contributions to the Hamiltonian, here we merely present a comprehensive sketch of the relevant processes and refer the reader to Ref.~\onlinecite{Ungar_Phonon-impact_2018} for the formal details.

\begin{figure}
	\centering
	\includegraphics{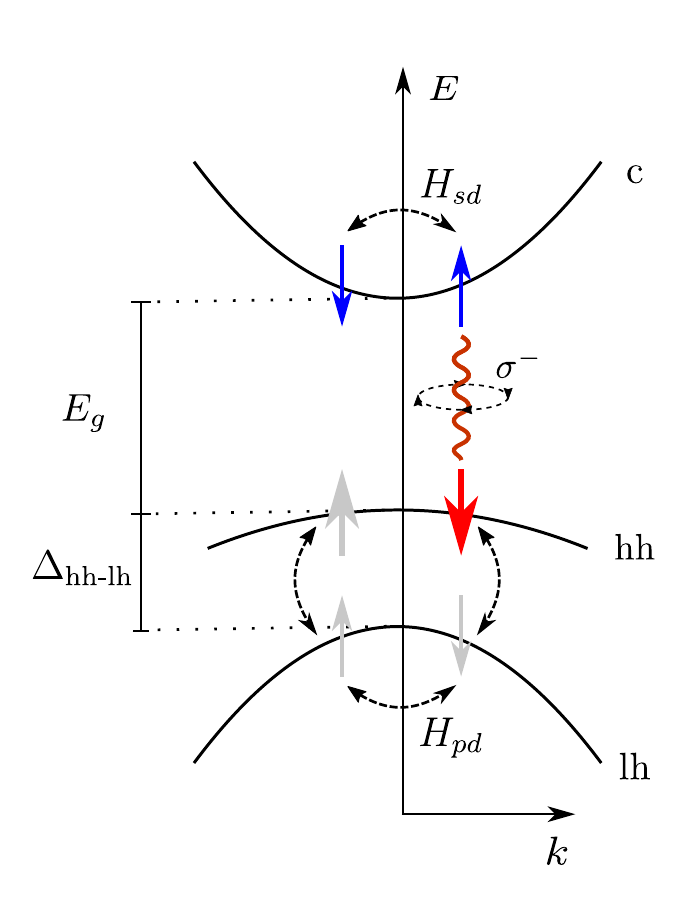}
	\caption{Sketch of the spin system under consideration. Illustrated are the dispersions of the conduction (c), the heavy-hole (hh), as well as the light-hole (lh) band in the parabolic effective-mass approximation.
The former two are separated by the band gap $E_g$, whereas the latter two are split by the hh-lh splitting $\Delta$.
Small arrows denote either spins in the conduction band with quantum number $s_z = \pm\frac{1}{2}$ or spins in the valence band with angular momentum quantum number $j_z = \pm\frac{1}{2}$, whereas the larger arrows correspond to hh spins characterized by $j_z = \pm\frac{3}{2}$.
The $sd$ exchange interaction ($H_{sd}$) mediates spin-flips in the conduction band and the $pd$ exchange interaction ($H_{pd}$) causes transitions between hh and lh spins.
Photons with $\sigma^-$ polarization (squiggly line) either create an electron-hh pair in the right spin channel or are emitted upon its recombination.
Grayed-out arrows indicate spin states which are not accessible after excitation of a hh with $j_z = -\frac{3}{2}$ due to energetic separation.}
	\label{fig:sketch}
\end{figure}

When discussing excitons in quantum wells, it is typically sufficient to account for the lowest $s$-like conduction band as well as the topmost $p$-like valence band, which are separated by the band gap $E_g$ \cite{Winkler_Spin-Orbit_2003}.
However, due to the confinement along the growth direction as well as strain in the semiconductor structure, the valence band splits in a heavy (hh) and a light hole (lh) branch separated by the hh-lh splitting $\Delta$, as sketched in Fig.~\ref{fig:sketch}.
There, small arrows denote electrons with spin quantum number $s_z = \pm\frac{1}{2}$ as well as lh states with total angular momentum quantum number $j_z = \pm\frac{1}{2}$.
Heavy holes with $j_z = \pm\frac{3}{2}$ are represented by thick arrows.
An optical excitation with $\sigma^-$-polarized light creates an electron-hole pair with $s_z = \frac{1}{2}$ and $j_z = -\frac{3}{2}$ in accordance with angular momentum conservation.
In Fig.~\ref{fig:sketch}, this process as well as the reverse process, where an electron-hole pair recombines and emits a photon, is represented on the right side of the figure by the squiggly line.

The previously mentioned carrier-impurity exchange interaction mediates spin-flips in this model, such that $H_{sd}$ couples spin-up and spin-down electrons in the conduction band and $H_{pd}$ couples lh spins in the valence band.
However, in the absence of mixing between heavy and light holes, $H_{pd}$ does not provide a direct coupling between the different hh spin states.
Thus, in order to flip its spin, the hh must pass through the lh states.
If the hh-lh splitting is large enough, this process is off-resonant on the order of $\Delta$ such that the hh spin is effectively pinned \cite{Uenoyama_Hole-relaxation_1990, Ferreira_Spin-flip_1991, Bastard_Spin-flip_1992, Crooker_Optical-spin_1997}.
Although there are mechanisms causing a hh-lh mixing such as the long-range part of the Coulomb interaction \cite{Maialle_Exciton-spin_1993, Maialle_Exciton-spin_1994}, the corresponding interaction energy is several orders of magnitude smaller than the typical energy of the carrier-impurity exchange interaction in DMSs \cite{Ungar_Trend-reversal_2018}, which is why we neglect the hh-lh mixing here.
In that case, the description of the spin dynamics can be effectively limited to only two exciton parabolas, namely that of an electron with spin up or down combined with a hh with $j_z = -\frac{3}{2}$.
These states are then conveniently labeled by the spin state of the electron since the hh spin does not change throughout the dynamics, as well as the corresponding center-of-mass wave number (or kinetic energy) of the exciton.

Treating the scattering of excitons with impurities, the optical excitation, as well as the exciton-phonon scattering as Markov processes, the equations of motion for the spin-up and spin-down exciton density $n_{\omega_1}^\ud$ at a fixed frequency $\omega_1$ can be written as \cite{Ungar_Phonon-impact_2018}
\begin{widetext}
\begin{align}
\label{eq:Markov-limit}
\ddt n_{\omega_1}^\ud =&\;
	\Gamma_E(\omega_1, t)
	- \Gamma_\t{rad}^\ud(\omega_1) \, n_{\omega_1}^\ud
	+ \Gamma_\t{sf}(\omega_1) (n_{\omega_1}^\du - n_{\omega_1}^\ud)
	\nn
	&+ \int_0^\infty d\omega \, D(\omega) \Lambda_{1s 1s}^{\omega_1 \omega} \bigg[ \Theta\big(\omega - \omega_1 - \omega_{\omega-\omega_1}^\t{ph}\big)
	\Big( n_{\omega}^\ud \big(1+n^\t{ph}(\omega - \omega_1)\big) - n_{\omega_1}^\ud n^\t{ph}(\omega - \omega_1) \Big)
	\nn
	&+ \Theta\big(\omega_1 - \omega - \omega_{\omega_1-\omega}^\t{ph}\big)
	\Big( n_{\omega}^\ud n^\t{ph}(\omega_1 - \omega) - n_{\omega_1}^\ud \big(1+n^\t{ph}(\omega_1 - \omega)\big) \Big)\bigg].
\end{align}
\end{widetext}
Considering a Gaussian pulse shape $E(t) = E_0\exp(-\frac{t^2}{2\sigma^2})$ with amplitude $E_0$, the optical generation of excitons is given by the rate
\begin{align}
\label{eq:optical-generation-rate}
\Gamma_E(\omega, t) &= \frac{1}{\hbar^2} E(t) E_0 |M_\ud|^2 \phi_{1s}^2 \int_{-\infty}^t d\tau \, e^{-\frac{\tau^2}{2\sigma^2}} \, \delta_{\omega,0}^b
\end{align}
with the matrix element $|M_\ud|^2$ containing the spin selection rules and $\phi_{1s} = R_{1s}(r = 0)$ denoting the radial part of the $1s$ exciton wave function evaluated at the origin.
The delta function $\delta_{\omega,0}^b$ reflects the fact that resonant optical excitation occurs only at the bottom of the exciton parabola, which is chosen as the origin of the energy scale here as indicated by the second subscript of the delta function.
To ensure a scalable as well as stable numerical evaluation of the delta function, we approximate it by a narrow Gaussian according to $\delta_{\omega,0}^b = \frac{1}{\sqrt{\pi} w_\t{b}}\exp(-(\hbar\omega/2w_\t{b})^2)$ that is normalized with respect to an integration over all positive frequencies.
Thus, we obtain an effectively broadened delta function as indicated by the superscript $b$.
The width of the Gaussian $w_\t{b} = 1\,$\textmu eV is chosen such that only states in close proximity of the exciton resonance at $\hbar\omega = 0$ couple to the light field.
We focus on excitation scenarios where few excitons compared to the number of impurities are excited such that the average impurity spin remains essentially constant during the dynamics.
Radiative decay is modeled via the spin-dependent rates $\Gamma_\t{rad}^\up(\omega) = \Gamma_0 \delta_{\omega,0}^b$ and $\Gamma_\t{rad}^\down(\omega) = 0$, where the latter reflects the optically dark nature of the spin-down exciton state.
As for the optical excitation, the slightly broadened delta function $\delta_{\omega,0}^b$ ensures that only states in the vicinity of the bottom of the exciton parabola can undergo radiative decay.

Without an external magnetic field, the spin-flip rate due to the scattering of excitons with the magnetic impurities in the crystal does not distinguish any spin orientation and only depends on the exciton frequency $\omega$.
It is given by \cite{Ungar_Quantum-kinetic_2017}
\begin{align}
\label{eq:spinflip-rate}
\Gamma_\t{sf}(\omega) = \frac{35 I \NMn M \Jsd^2}{12 \hbar^3 V d} F_{\etah 1s 1s}^{\etah \omega \omega}
\end{align}
with a factor $I = 3/2$ from the projection of the wave function onto the quantum well whose thickness is given by $d$.
The number of Mn impurities in the sample is given by $\NMn$ and $M$ is the exciton mass.
Furthermore, $F_{\etah 1s 1s}^{\etah \omega \omega}$ denotes the frequency-dependent exciton form factor which appears due to the projection of the dynamics onto the exciton basis and which contains the exciton wave function \cite{Ungar_Quantum-kinetic_2017}.
An explicit expression of the form factor can be found in the Appendix of Ref.~\onlinecite{Ungar_Phonon-induced_2019}.

The integral that appears in Eq.~\eqref{eq:Markov-limit} contains expressions stemming from the exciton-phonon scattering.
There, $D(\omega) = \frac{VM}{2\pi\hbar d}$ is the constant density of states for a two dimensional system and $\Theta(\omega)$ is the Heaviside step function and the phonon density is assumed to follow a thermal occupation according to $n^\t{ph}(\omega) = 1/(\exp(\hbar\omega/k_BT) - 1)$.
Finally, $\Lambda_{1s 1s}^{\omega_1 \omega}$ is the exciton-phonon matrix element which contains the influence of the exciton wave function and the exciton-phonon coupling.
For an explicit expression of this matrix element, the reader is referred to the Appendix of Ref.~\onlinecite{Ungar_Phonon-induced_2019}.
Note that the $z$ component of the spin can be extracted from the spin-up and spin-down exciton density via $s_\omega^z = \frac{1}{2} (n_\omega^\up - n_\omega^\down)$.

However, as shown in previous theoretical works for excitons \cite{Ungar_Quantum-kinetic_2017, Ungar_Trend-reversal_2018, Ungar_Many-body_2018, Ungar_Phonon-impact_2018}
as well as quasifree conduction-band electrons \cite{Thurn_Non-Markovian_2013, Cygorek_Non-Markovian_2015, Cygorek_Nonperturbative-correlation_2016}, a Markovian treatment of the typically dominant carrier-impurity exchange interaction such as given by Eq.~\eqref{eq:Markov-limit} is often insufficient since it cannot capture correlation effects beyond the single-particle picture.
Furthermore, typical theoretical descriptions of the spin dynamics in DMSs based on Fermi's golden rule \cite{Nawrocki_Exchange-Induced_1981, Bastard_Spin-flip_1992, 
Camilleri_Electron-and_2001, Cywinski_Ultrafast-demagnetization_2007, Tsitsishvili_Magnetic-field_2006} artificially enforce momentum as well as energy conservation on the single-particle level, where the former assumption is violated for systems with few randomly localized scatterers such as DMSs \cite{Cygorek_Influence-of_2017} and the latter neglects the energy-time uncertainty.
To account for these effects we have developed a full quantum kinetic theory which explicitly keeps exciton-impurity correlations as dynamical variables, which also allows us to straightforwardly describe scattering processes that do not conserve momentum \cite{Thurn_Quantum-kinetic_2012, Ungar_Quantum-kinetic_2017}.
The QKT has also just recently been extended to account for phonon scattering on the Markov level in Ref.~\onlinecite{Ungar_Phonon-impact_2018}, where also the complete equations of motion can be found.
In the present article, we will make use of this advanced theory to compare it with the more intuitive Markovian theory presented above.
For this purpose, radiative decay is also included in the QKT in a similar manner to Eq.~\eqref{eq:Markov-limit} (cf. also the Supplemental Material of Ref.~\onlinecite{Ungar_Many-body_2018}).

\subsection{Minimal model for overshooting behavior}
\label{subsec:Minimal-model-for-overshooting-behavior}

To analyze the requirements of observing an overshooting behavior in some physical quantity, we consider the following system of coupled differential equations for the time-dependent quantities $a$ and $b$:
\begin{subequations}
\label{eqs:a-b-coupled}
\begin{align}
\ddt a &= - \kappa_a a + \lambda (b - a)
	\\
\ddt b &= - \kappa_b b + \lambda (a - b)
\end{align}
\end{subequations}
The model includes decay rates $\kappa_a$ and $\kappa_b$ for $a$ and $b$, respectively, and allows for a transfer between the two quantities via the rate $\lambda$.
Since the timescales of this model are solely determined by the value of the rates, we rescale the time such that it becomes dimensionless.
The signs in Eqs.~\eqref{eqs:a-b-coupled} are chosen such that, for typically encountered physical systems, $\kappa_{a/b} \geq 0$ and $\lambda \geq 0$.
Although the transfer rate in general may be different for a transfer from $a$ to $b$ compared with one in the opposite direction, we limit the discussion to equal transfer rates since this is also the case for the spin-flip scattering in Eq.~\eqref{eq:Markov-limit}, where the scattering rate is given by Eq.~\eqref{eq:spinflip-rate} and does not depend on the scattering direction.
The main difference with respect to the full model given by Eq.~\eqref{eq:Markov-limit} is that the phonon scattering is completely left out.
In terms of Eqs.~\eqref{eqs:a-b-coupled}, such a contribution would cause the quantities $a$ and $b$ to become a continuum of values which are all coupled to one another.
However, in order to understand to origin of the overshooting behavior, the phonon contribution can be neglected, especially at low temperatures where phonon absorption is limited and phonon emission cannot occur for optically generated excitons with vanishing center-of-mass momentum.

\begin{figure}
	\centering
	\includegraphics{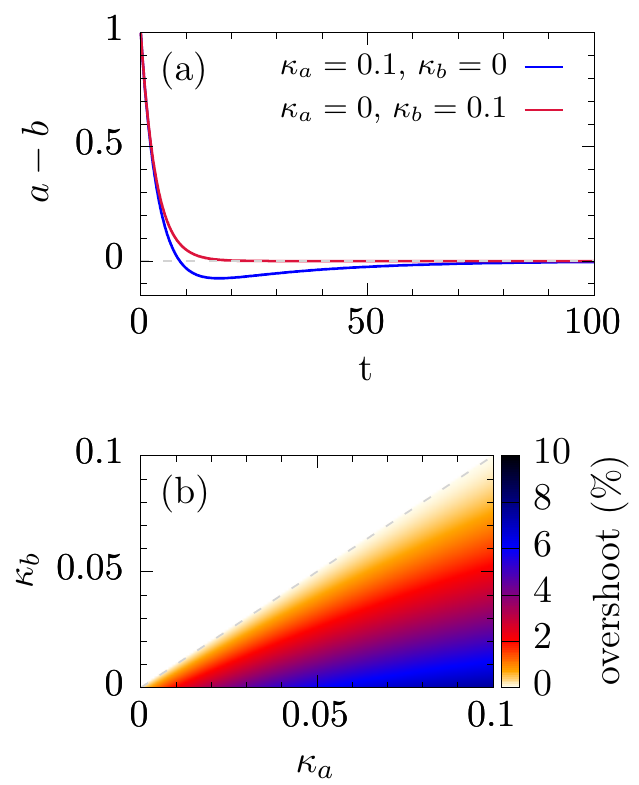}
	\caption{Dynamics of the minimal model given by Eqs.~\eqref{eqs:a-b-coupled} with $\lambda = 0.1$.
We show (a) the dynamics of $a-b$ for two representative choices of $\kappa_a$ and $\kappa_b$ as well as (b) a color-coded map of the parameter space where an overshoot occurs.
The percentage value of the overshoot refers to how much $a-b$ maximally dips below zero with respect to its initial value.}
	\label{fig:model}
\end{figure}

Coming back to the minimal model at hand and using the initial conditions $a(0) = 1$ and $b(0) = 0$, Eqs.~\eqref{eqs:a-b-coupled} can be solved exactly and $a$ and $b$ are both strictly between zero and one.
However, instead of solving for the dynamics of $a$ and $b$ separately, we are interested in a third variable which is represented by a linear combination of the two.
Supposing that, e.g., $a$ and $b$ are spin channels corresponding to spin up and spin down, respectively, the total spin in the system is $s \sim a - b$.
Depending on the constants of the model, such a variable may then display the overshooting behavior we are looking for since the analytic solution will in general contain a biexponential decay.

Instead of discussing the analytic solution, which is lengthy and does not clearly display the physical insights we seek, we plot the resulting dynamics for $a-b$ for two representative choices of parameters in Fig.~\ref{fig:model}(a).
Indeed, choosing $\kappa_a = 0.1$ and $\kappa_b = 0$, an overshoot appears which results from the faster decay of $a$ compared with $b$, such that $a-b$ takes on negative values after a certain point in the dynamics.
Reversing the situation by choosing $\kappa_a = 0$ and $\kappa_b = 0.1$ then leads to no overshooting behavior since $b < a$ throughout the dynamics.
Furthermore, due to the transitions between $a$ and $b$, as soon as either $\kappa_a$ or $\kappa_b$ are finite both variables eventually decay to zero, which thus also holds for $a-b$ as can be seen in the figure.

To obtain an overview of the parameter space where an overshoot occurs and how large it can actually be, we plot the overshoot versus both decay rates, $\kappa_a$ and $\kappa_b$, in Fig.~\ref{fig:model}(b).
Any given value for the overshoot reflects the percentage of how much $a-b$ maximally dips below zero compared with its initial value of $(a-b)|_{t=0} = 1$, i.e., we define the overshoot as the distance from the global minimum of the curve with respect to its long-time value, which is zero in this case (also indicated by the gray dashed line in Fig.~\ref{fig:model}).
As motivated above, the figure confirms that an overshoot can only occur in the region where $\kappa_b < \kappa_a$ below the gray dashed line and is most pronounced when $\kappa_b = 0$.
The maximum overshoot obtained in the latter case is slightly below $10\%$.
Although this model is strongly simplified, it nevertheless allows for an intuitive understanding of the physical processes encountered in the exciton spin dynamics.

Concerning DMSs, the minimal model can be applied as follows.
Taking a look at Fig.~\ref{fig:sketch}, it is apparent that an optical pump pulse promotes an electron from the valence band to the conduction band.
However, due to the optical selection rules, this creates an exciton consisting of a hh with $j_z = -\frac{3}{2}$ and an electron with $s_z = \frac{1}{2}$.
Since directly after the pulse only this state is occupied, it corresponds to channel $a$ in the minimal model, which has a finite occupation at $t = 0$.
If the hh-lh splitting is large enough, only spin flips in the conduction band are likely to occur since the hh-lh splitting acts as an energy barrier which prevents a spin flip of the hh, effectively pinning it along its initial orientation.
After the spin flip of an electron mediated by $H_{sd}$, one ends up with an exciton consisting again of a hh with $j_z = -\frac{3}{2}$ and an electron with $s_z = -\frac{1}{2}$, which then corresponds to channel $b$ in the minimal model.
In general, any excitation can leave the system only via radiative decay, corresponding to $\kappa_a$ and $\kappa_b$ for excitations involving a spin-up and a spin-down electron, respectively.
Since the optical selection rules also apply in the case of radiative decay, only the population of channel $a$ can decay since channel $b$ is optically dark.
For the minimal model, this means that one is in the regime where only $\kappa_a$ has a finite value.
As long as only excitons with vanishing center-of-mass momentum are present in the system, which is true in the absence of phonon scattering, the transfer rate $\lambda$ between the two channels $a$ and $b$ can be exactly identified with $\Gamma_\t{sf}(0)$ as given by Eq.~\eqref{eq:spinflip-rate}.

\section{Simulations of exciton-spin overshoots}
\label{sec:Simulations-of-exciton-spin-overshoots}

In this section, simulations of the exciton spin dynamics in DMSs under the influence of exciton-impurity as well exciton-phonon scattering are performed while also accounting for radiative decay.
Special emphasis is put on the observation of spin overshoots as well as a comparison of two different theoretical approaches, namely the MT and the more advanced QKT.
First, the appearance of overshoots is investigated using both theories and the impact of higher temperatures and, thus, stronger phonon scattering is discussed.
Second, we suggest specific parameter studies which would allow for an experimental determination of the origin of the spin overshoot in a given sample.
For all calculations, we model a $15\,$nm wide Zn$_{1-x}$Mn$_{x}$Se quantum well with doping fraction $x$ and suppose a resonant excitation of the $1s$-$hh$ exciton with a $100\,$fs long pulse.
For the radiative decay we assume a rate of $0.1\,$ps$^{-1}$, such that the resulting lifetime is in line with typical experiments \cite{Kalt_Optical-and_1992, Runge_Relaxation-kinetics_1995, Chen_Spin-dynamics_2003, Poltavtsev_Damping-of_2017}.
The values for the coupling constants as well as all other material parameters are the same as in Ref.~\onlinecite{Ungar_Phonon-impact_2018}.

\subsection{Phonon influence on spin overshoots}
\label{subsec:Phonon-influence-on-spin-overshoots}

\begin{figure*}
	\centering
	\includegraphics{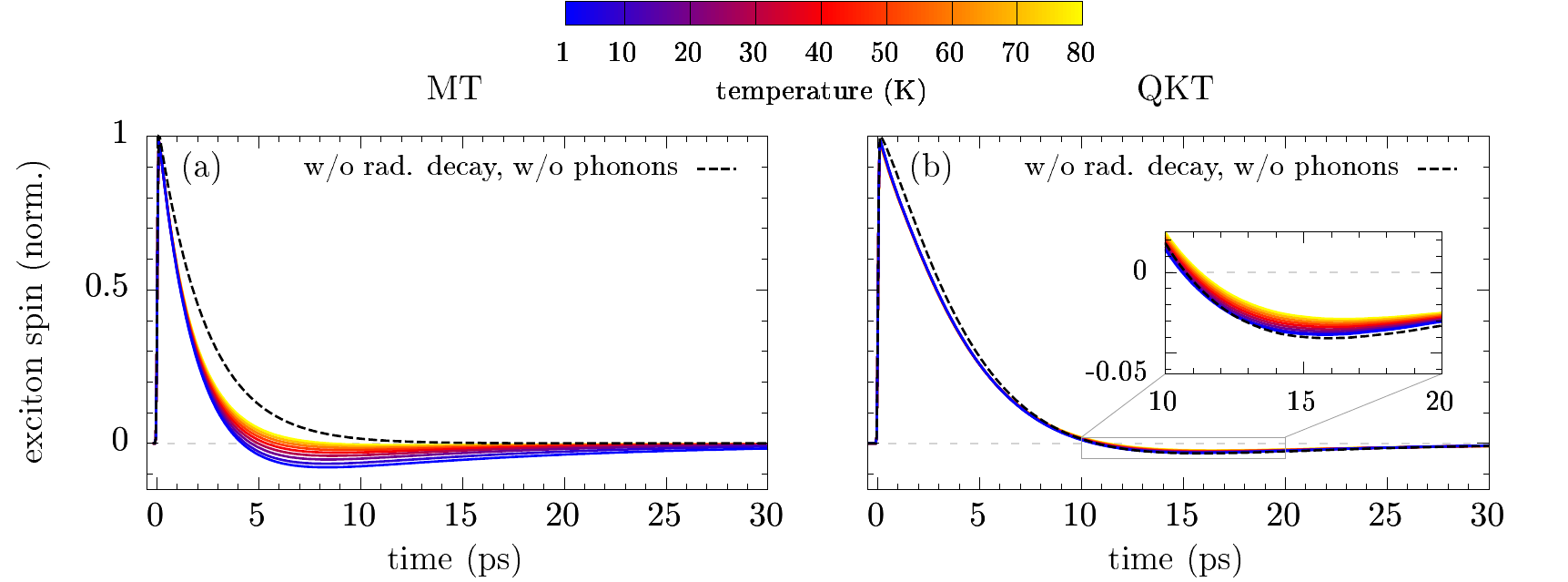}
	\caption{Exciton-spin dynamics normalized to the maximum value after the optical excitation pulse for different temperatures for a Zn$_{0.975}$Mn$_{0.025}$Se quantum well.
We compare (a) results using a completely Markovian theory (MT) with (b) simulations obtained by a quantum kinetic treatment (QKT) of the exciton-impurity scattering.
In addition, simulation results without radiative decay and without phonon scattering are provided for both theoretical approaches.
The inset in panel (b) shows a magnified view of the area indicated by the gray rectangle.}
	\label{fig:overshoot}
\end{figure*}

Without an external magnetic field, optically oriented exciton spins are expected to decay to a vanishing net spin polarization after a characteristic time given by the inverse of the rate in Eq.~\eqref{eq:spinflip-rate}.
The typically resulting dynamics for the first $30\,$ps is shown in Fig.~\ref{fig:overshoot}(a), where the data calculated using the MT given by Eq.~\eqref{eq:Markov-limit} is normalized with respect to the maximum spin polarization after the pulse.
Without radiative decay and the phonon influence, Eq.~\eqref{eq:Markov-limit} reduces to just the optical excitation and the spin-decay rate given by Eq.~\eqref{eq:spinflip-rate} so that a monoexponential decay to zero occurs after the optical orientation (cf. black dashed line).
With radiative decay and phonon scattering, however, this behavior changes drastically:
the MT now predicts an overshoot of the exciton spin that becomes less pronounced with increasing temperature.
For very low temperatures on the order of $1\,$K the overshoot is almost $10\%$, whereas it vanishes almost completely for $80\,$K.

To explain this behavior we turn to the minimal model introduced in Sec.~\ref{subsec:Minimal-model-for-overshooting-behavior} in Eqs.~\eqref{eqs:a-b-coupled}.
There, the overshoot was traced back to a slower decay of the $b$ component compared with the $a$ component in a signal given by $a-b$.
Consequently, $a-b$ then follows a biexponential decay and is thus able to dip below zero before it completely decays, thus causing an overshoot of the signal.
As already motivated in this discussion, identifying $a$ with the spin-up and $b$ with the spin-down exciton density allows us to straightforwardly explain the curves in Fig.~\ref{fig:overshoot}(a) by comparing them to the results shown in Fig.~\ref{fig:model}.
In the present case, the decay rate of the spin-up channel is given by $0.1\,$ps$^{-1}$, whereas the spin-down channel is unaffected by the decay since it represents a dark exciton.
For comparison, the spin-decay rate (which corresponds to the transfer rate $\lambda$ in the minimal model) is roughly $0.44\,$ps$^{-1}$ for the parameters in Fig~\ref{fig:overshoot}.
Since we also find an overshoot of about $10\%$ in Fig.~\ref{fig:model} for the appropriate parameters, we can conclude that the overshoot observed in Fig.~\ref{fig:overshoot}(a) for very low temperatures is entirely due to radiative decay.

To understand the influence of phonons on the overshooting behavior one has to be aware that phonon absorption and emission processes are not equally likely for low temperatures
and only become similar in probability when the temperature is high enough.
Since phonon absorption obviously requires the presence of phonons in the system and is proportional to $n^\t{ph}$ in Eq.~\eqref{eq:Markov-limit}, only phonon emission proportional to $1+n^\t{ph}$ can occur in the low-temperature limit.
However, keeping in mind that excitons are optically created with nearly vanishing center-of-mass momenta, there are simply no states with lower kinetic energies for excitons available to scatter to such that a phonon could be emitted.
This explains why for temperatures on the order of a few K the dynamics is virtually unaffected by phonons since neither absorption nor emission are likely to occur.
Figure~\ref{fig:overshoot}(a) also shows that the magnitude of the overshoot decreases for elevated temperatures, which can be explained by the fact that the now more probable phonon absorption increases the exciton kinetic energy and, thus, shifts the center-of-mass momenta away from zero.
But as light only couples to excitons with nearly vanishing momenta, excitons with larger momenta are optically dark and, thus, are no longer affected by radiative decay.

Turning now to Fig.~\ref{fig:overshoot}(b), where simulations for the same parameters are shown using the QKT, we see a strikingly different behavior.
There, even a calculation without radiative decay and no phonon scattering leads to an overshoot, albeit with a smaller magnitude compared with the corresponding predictions of the MT.
As pointed out in a previous publication \cite{Ungar_Quantum-kinetic_2017}, the overshoot in the QKT without radiative decay is an effect that cannot be reproduced on the Markov level [cf. black dashed line in Fig.~\ref{fig:overshoot}(a)] since it requires exciton-impurity correlations that are not captured in an effective single-particle theory.
In fact, the overshoot is related to the behavior of the memory kernel which is given by a sinc-like function that shows decaying oscillations  \cite{Cygorek_Non-Markovian_2015}.
Since the frequency of these oscillations depends on the energy of the carriers, the oscillations typically become averaged out when a distribution of carriers is considered so that only an overshoot remains.
In general, this quantum kinetic effect becomes more pronounced in nanostructures compared with bulk systems \cite{Thurn_Non-Markovian_2013}.
Note also the slower decay of the exciton spin when using the QKT compared with the MT which can be traced back to a cutoff of the memory kernel at the bottom of the exciton parabola \cite{Ungar_Quantum-kinetic_2017}.
Since the MT assumes a vanishing memory, it is unable to account for either of these effects.
Apart from the fact that the QKT predicts an overshoot even without radiative decay, Fig.~\ref{fig:overshoot}(b) also reveals that phonons have basically no impact on the spin dynamics for resonantly excited excitons up to the maximum temperature of $80\,$K considered here \cite{Ungar_Phonon-impact_2018}.
This means that, for typical DMSs, correlation effects dominate and suppress overshoots due to radiative decay.

It is important to emphasize that the Markovian results shown in Fig.~\ref{fig:overshoot}(a) become valid in non-DMS systems that are not dominated by strong correlations due to the exciton-impurity interaction.
In nonmagnetic semiconductors, provided there is a suitably strong spin-flip mechanism such as, e.g., the Dyakonov-Perel' mechanism due to spin-orbit coupling \cite{Wu_Spin-dynamics_2010, Ungar_Ultrafast-spin_2015, Cosacchi_Nonexponential-spin_2017}, one should therefore indeed expect overshoots because of radiative decay.
For the quantum kinetic results shown in Fig.~\ref{fig:overshoot}(b), however, exciton-impurity correlations are the crucial ingredient for the overshooting behavior.
Indeed, this is typical for DMSs since the many-body interaction is strongly boosted by the magnitude of the coupling constants $\Jsd$ and $\Jpd$ in these systems \cite{Ungar_Many-body_2018}.
From this point of view, the results obtained by the MT may be interpreted as corresponding to materials where radiative decay dominates.
The range of validity of the QKT of course includes that of the MT.
Thus, deviations between these two levels of theory, as found for DMSs, indicate a dominance of many-body physics.

\begin{figure}
	\centering
	\includegraphics{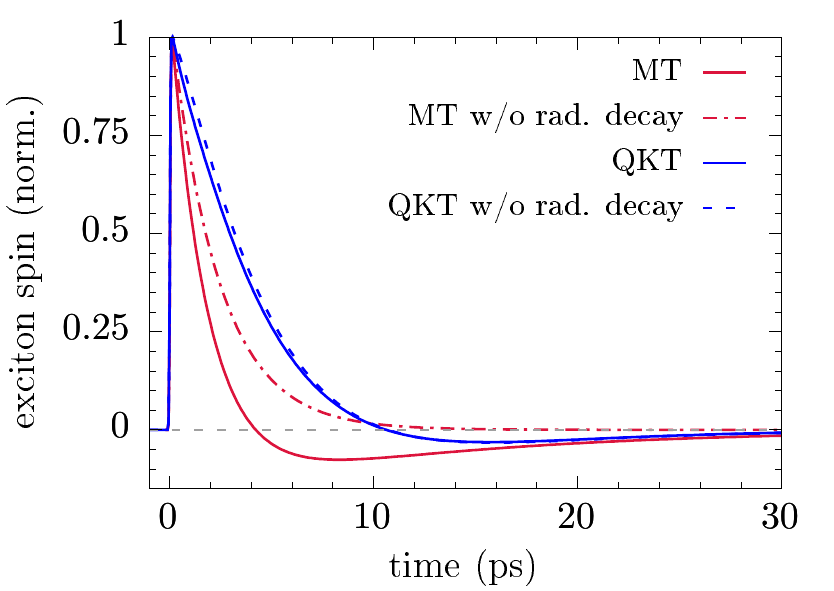}
	\caption{Direct comparison of the overshoot predicted by the Markovian theory (MT) and the quantum kinetic theory (QKT) with and without radiative decay.
The simulations are performed for a Zn$_{0.975}$Mn$_{0.025}$Se quantum well at $4\,$K including phonon scattering.}
	\label{fig:comp}
\end{figure}

A more direct comparison between the results of the two theories regarding the influence of radiative decay is presented in Fig.~\ref{fig:comp} for a Zn$_{0.975}$Mn$_{0.025}$Se quantum well at $4\,$K.
First of all, not only does the overshoot in the MT appear sooner, but it is also more pronounced compared with the QKT result.
However, as soon as radiative decay is switched off, the MT reverts to a monoexponential decay in the manner of Fermi's golden rule and no longer displays an overshooting behavior.
In contrast, the influence of radiative decay on the QKT simulations is very limited.
Although it does cause a slightly faster decay for the first few picoseconds, after approximately $10\,$ps its effect is completely negligible.
This again shows that the spin dynamics in DMSs is dominated by correlations induced by the exciton-impurity exchange interaction.

\subsection{Markovian vs. quantum kinetic predictions}
\label{subsec:Markovian-vs-quantum-kinetic-predictions}

Regarding experiments, it is an important question to ask how the origin of a spin overshoot can be determined, i.e., how one can decide whether it is dominated by radiative decay or many-body correlations.
To answer this question, we compare results of the MT with those of the QKT while varying either the temperature or the doping fraction of the DMS quantum well.

\begin{figure}
	\centering
	\includegraphics{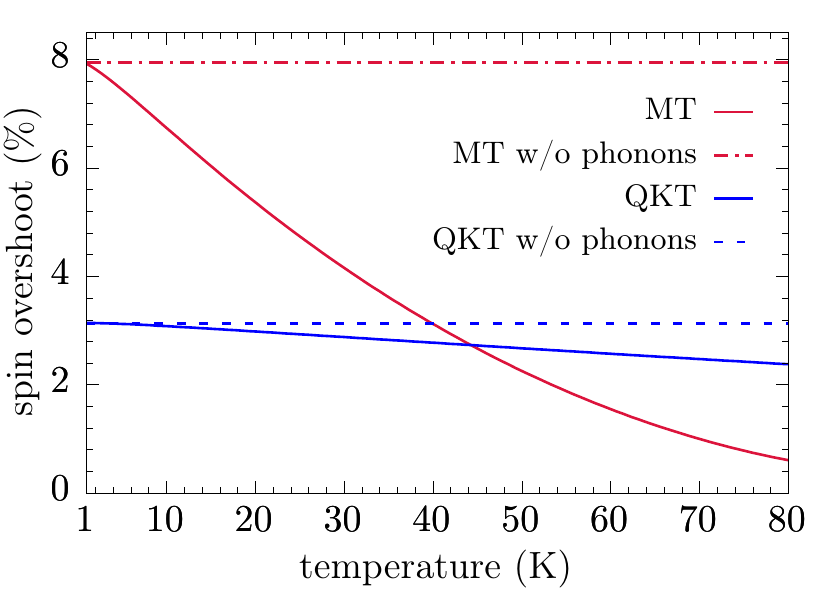}
	\caption{Overshoot of the exciton spin polarization with respect to the maximum polarization reached after optical excitation as a function of the temperature for a Zn$_{0.975}$Mn$_{0.025}$Se quantum well.
We compare simulations using the Markovian theory (MT) with results obtained by the quantum-kinetic approach (QKT).
Results without carrier-phonon interaction (w/o phonons) are also shown.}
	\label{fig:min_value_vs_T}
\end{figure}

As can already be seen from Fig.~\ref{fig:overshoot}, the influence of phonons and, thus, changing the temperature is quite different for the MT compared with the QKT.
To make this different behavior more apparent, Fig.~\ref{fig:min_value_vs_T} displays the spin overshoot discussed in the previous section as a function of the temperature for the two theoretical approaches with and without phonons, respectively.
In the MT, increasing the temperature causes a steep drop of the spin overshoot from $8\%$ to almost zero when phonons are accounted for.
In contrast, while a phonon influence is visible in the QKT, it does not significantly affect the spin overshoot and only causes it to decrease from about $3\%$ at $0\,$K to $2.5\%$ at $80\,$K.
The figure also shows that the spin overshoot at low temperatures is generally smaller in the QKT compared with the MT.

\begin{figure}
	\centering
	\includegraphics{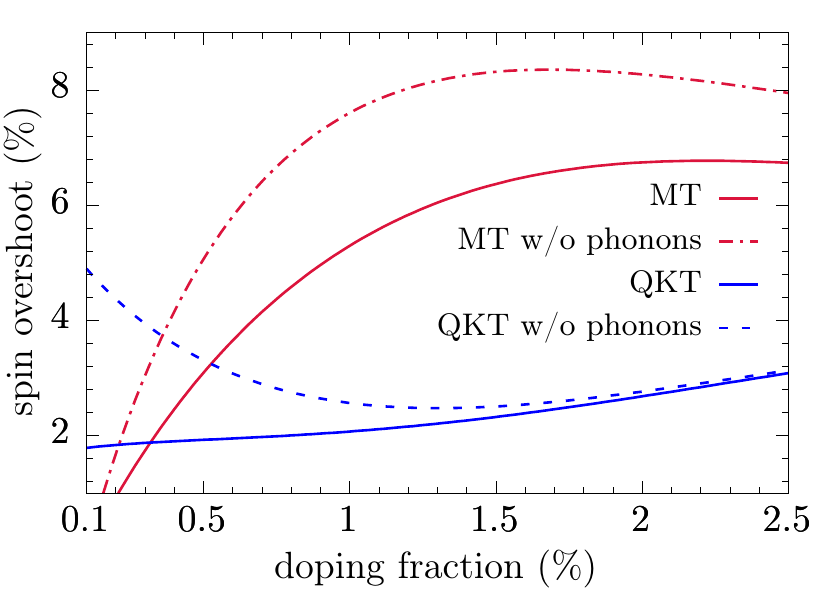}
	\caption{Overshoot of the exciton spin polarization with respect to the maximum polarization reached after optical excitation as a function of the doping fraction for a Zn$_{1-x}$Mn$_{x}$Se quantum well at a temperature of $10\,$K.
We compare simulations using the Markovian theory (MT) with results obtained by the quantum-kinetic approach (QKT).
Results without carrier-phonon interaction (w/o phonons) are also shown.}
	\label{fig:min_value_vs_xMn}
\end{figure}

Similarly, the theories predict a different dependence of the spin overshoot on the doping fraction of the quantum well, as shown in Fig.~\ref{fig:min_value_vs_xMn}.
In the MT, the doping fraction basically scales the spin-decay rate given by Eq.~\eqref{eq:spinflip-rate} since it appears as a prefactor there.
This means that for fewer Mn ions the spin-decay rate becomes significantly smaller than the constant radiative decay rate, which in turn causes a much faster decrease of the spin-up population compared with the scattering to the spin-down state.
Thus, the majority of excitons have already decayed before a significant spin-down population can be reached and only a small overshoot occurs.
Increasing the doping fraction increases the spin-flip scattering and thus allows for a more pronounced overshoot that begins to slightly decrease again for doping fraction of about $2\%$ and larger.
In that case, the spin-flip scattering rate given by Eq.~\eqref{eq:spinflip-rate} is at least four times larger than the radiative decay rate, thus making spin-flips very efficient so that the imbalance due to radiative decay becomes less pronounced.
As seen before, phonons generally decrease the spin overshoot in this model.

Turning to the results of the QKT, we find an overall smaller overshoot which is, however, enhanced compared with the MT at low doping, where also the phonon influence is most noticeable.
Without phonons, the spin overshoot continuously increases with increasing impurity content because of the similarly increasing correlation energy, which is roughly proportional to the doping fraction \cite{Ungar_Many-body_2018}.
The decreasing impact of phonons on the spin overshoot with rising impurity content is explained by the quantum kinetic redistribution of excitons in $K$ space which is not captured on the Markov level.
This redistribution is made possible by the finite exciton-impurity correlations that cause the proper many-body eigenstates of the system to be a combination of states with different center-of-mass momenta, thus effectively smearing out the exciton population.
This additional scattering towards higher momenta can be substantially larger than the phonon scattering, especially at high doping fractions and on short time scales \cite{Ungar_Phonon-impact_2018}.
All in all, it becomes clear that an overshoot stemming from carrier-impurity correlations is much more robust against variations of the temperature as well as a change of the doping fraction compared with the effect of radiative decay.

\section{Conclusion}
\label{sec:Conclusion}

We have investigated the origin of overshoots appearing in the spin dynamics of resonantly excited excitons in DMS quantum wells.
On the Markov level, overshoots can appear as a consequence of radiative decay combined with strong spin-flip scattering mechanisms, such as the carrier-impurity exchange interaction in DMSs.
Using a minimal model where only radiative decay and a spin-flip rate between two spin populations are present even allowed for a straightforward estimation of the parameter space where overshoots can occur.

Modeling realistic samples, we have investigated the temperature influence on spin overshoots by accounting for LA phonon scattering within a Markovian description that is expected to be valid for situations where no strong many-body correlations are built up (e.g. in nonmagnetic semiconductors with spin flips resulting from spin-orbit coupling).
Phonons reduce the spin overshoot and can even cause it to nearly vanish close to liquid nitrogen temperatures.
The reason for this is the enhanced scattering of excitons away from the region near $K = 0$ towards higher center-of-mass momenta and, thus, optically dark states.
If the temperature is low enough, however, overshoots of up to $10\%$ are predicted by our model.
To the best of our knowledge, such a nonmonotonic spin decay has so far only been found using a more advanced quantum kinetic theory \cite{Cygorek_Non-Markovian_2015, Ungar_Quantum-kinetic_2017}.

Having found spin overshoots on the Markov level caused by radiative decay obviously raises the question as to how one can determine the origin of an overshooting behavior observed in experiments.
We provide an answer to this question by directly comparing parameter dependences of the spin overshoot as obtained by the MT as well as the QKT.
When the exciton-impurity interaction is treated quantum kinetically, a spin overshoot appears as a consequence of the many-body nature of the system even without radiative decay and correlations beyond the Markov level are required to obtain this behavior.
Comparing results of the MT and the QKT reveals that the overshoot in the QKT is much more stable against the phonon influence, albeit it is not as pronounced as in the MT at low temperatures.
Furthermore, the two theories predict a different dependence of the magnitude of the spin overshoot on the doping fraction.
All in all, our theoretical investigations reveal that radiative decay has in fact little to no impact on the exciton spin dynamics in DMSs.
Instead, the dynamics is completely dominated by correlations caused by the exciton-impurity exchange interaction.

Although we focus on the spin dynamics in DMSs, the results of our paper are not restricted to this specific material system.
Instead, they can be used to analyze the exciton spin dynamics also in standard nonmagnetic semiconductors where radiative decay combined with a spin-flip mechanism plays a role.
Since non-DMS samples can be expected to be much less affected by the carrier-impurity correlations appearing in DMSs, we expect the results of the Markov approximation to be relevant in this case.
In this sense, our study provides a means to trace back the origin of an observed spin overshoot since it allows one to discriminate between overshoots dominantly caused by either radiative decay or non-Markovian effects.

\section{Acknowledgements}
\label{sec:Acknowledgements}

We gratefully acknowledge the financial support of the Deutsche Forschungsgemeinschaft (DFG) through Grant No. AX17/10-1.

\bibliography{references}
\end{document}